# Controlled electrochemical functionalization of CNT fibers: structure-chemistry relations and application in current collector-free all-solid supercapacitors.

*Evgeny Senokos[a,b,c, §], Moumita Rana[a, §], Cleis Santos[a], Rebeca Marcilla[b]\* and Juan J. Vilatela[a]\*.*

[a] IMDEA Materials Institute, c/ Eric Kandel 2, Getafe 28906, Madrid, Spain

[b] IMDEA Energy Institute, Parque Tecnológico de Móstoles, Avda. De la Sagra 3, 28935 Móstoles, Madrid, Spain

[c] E. T. S. de Ingenieros de Caminos, Universidad Politécnica de Madrid, 28040 Madrid, Spain

[§] These authors contributed equally to this work

\* rebeca.marcilla@imdea.org, juanjose.vilatela@imdea.org

ABSTRACT

Chemical functionalization of nanocarbons is an important strategy to produce electrochemical systems with higher energy/power density by generating surface functional groups with additional faradaic contribution, by increasing their surface area and correspondent capacitive contribution and by improving compatibility with aqueous electrolytes and other active materials, such as pseudocapacitive metal-oxides. Here we present an electrochemical method to simultaneously swell and functionalize large electrodes consisting of fabrics of macroscopic fibers of carbon nanotubes that renders the material hydrophilic and produces a substantial increase of specific capacitance and energy density in aqueous electrolytes. Through in-depth characterization of the carbon nanotube fibres (CNTF) by Raman spectroscopy, transmission electron microscopy, X-ray photoelectrocn spectroscopy (XPS) and small-angle X-ray scattering (SAXS) we identify various contributions to such improvements, including surface oxidation, tubular unzipping, debundling and inter-bundle swelling. Changes in hydrophilicity of functionalized CNTF are determined by analyzing the dynamics of spreading of polar and nonpolar liquids in the electrode. The extracted contact angles and polar and dispersive surface energy components for different treatment conditions are in agreement with changes in dipole-moment obtained by XPS. Finally, functionalized CNTF electrodes were employed in current collector-free solid flexible supercapacitors, which show enhanced electrochemical properties compared to as-produced hydrophobic ones.

1. INTRODUCTION

The superlative properties of nanocarbons continue to fuel the interest in macroscopic architectures that efficiently exploit their "molecular" properties. The exceptional stiffness, charge mobility and electrochemical stability of CNTs, for example, makes them attractive for applications ranging from lightweight composites, to electrochemical charge storage/transfer processes to biomedical applications.[1,2,3] In this quest, macroscopic CNT fibres, yarns and fabrics have emerged as attractive systems in which the CNTs associate in long coherent bundle domains that favor inter-tube charge and stress transfer, while also leaving large mesoporous gaps between bundles and thus giving rise to a large porosity. Such structure leads to an unusual combination of bulk mechanical toughness, electrical conductivity and electrochemical stability above that of many metals, combined with a large specific surface area above 250 $m^2/g$. These property envelop is partly the reason of the increasing use of CNT fibres as electrodes/current collectors in multiple energy storage, transfer and conversion devices, including those requiring light weight and augmented mechanical properties[4].

In this context, it is of interest to interface the porous CNT fibre with materials that have poor chemical affinity with graphitic surfaces. Typical examples are aqueous electrolytes, aqueous precursor solutions for subsequent hybridization (e.g. hydrothermal, sol-gel), biological media, and polar solids.[5]

In the case of nanocarbons in powder form, surface functionalization is conveniently carried out by wet chemical methods, such as the well-known oxidation in acid media, diazo-coupling or nitrene cycloaddition or intercalation by small organic groups, depending on its purpose.[6] The introduction of polar groups on the CNT surface enables spontaneous wetting by polar reactants and solvents, particularly water, which is otherwise limited by the extreme hydrophobicity of CNTs. It can also modify effective surface area, capacitance, catalytic activity, electrical conductivity, Seebeck coefficient, amongst other properties.

In contrast, the aggregated porous structure of CNT fibres does not lend itself to the use of traditional wet-chemical methods of functionalisation. Inhomogenous functionalization as a result of diffusion-limited reactions into the porous material is an example of one of the inherent difficulties identified. In view of this challenge, gas-phase oxidative functionalization methods would seem attractive. We recently developed a protocol using ozone and studied the effect of functionalization conditions on surface chemistry and electrochemical properties, including qualitatively discriminating faradaic, geometric and quantum capacitance contributions by using aqueous and ionic liquid electrolytes.[7] Although successful, this functionalization method presents limitations to produce uniform treatments in samples with non-planar shape or large areas, imposed by the configuration of the ozone generator lamp, the need for a closed chamber and a compositional gradient arising from diffusion of gas molecules.

Previous studies demonstrated that electro-oxidation methods are an interesting alternative to produce pseudocapacitance-containing activated carbons with enhanced energy storage capabilities.[8] With the aim of extending the set of chemical methods to modify CNT fibres and improve our understanding of the relation between surface chemistry and electrochemical properties, in this work we report a systematic study on the functionalization of CNT fiber assemblies by an electrochemical method in an aqueous electrolyte. Electrochemical functionalization is appealing as it can offer uniform diffusion of reactants and thereby homogeneous functionalization of conducting materials. It is also interesting to consider the possibility to electrochemically partially exfoliate graphitic layers, and produce a corresponding increase in surface area and associated properties.[9]

The effect of electrical bias on physiochemical properties of CNT fibres was investigated by Raman spectroscopy, transmission electron microscopy (TEM), X-ray photoelectron spectroscopy (XPS) and synchrotron small-angle X-ray scattering (SAXS). Combined with experiments on wicking of polar as well as nonpolar liquids in functionalized CNT fibers, we could establish a useful relation between treatment parameters, surface chemistry, pore structure and bulk wetting properties. The electrochemical properties of homogeneously functionalized CNT fibers were found to be superior with respect to pristine CNT fibers, showing an enhancement of over 4 times in specific capacitance in aqueous electrolyte (60 F g$^{-1}$ *vs*. 10 F g$^{-1}$) due to the presence of newly-acquired redox active functional groups and hydrophilicity of functionalized CNT fibers. All-solid flexible supercapacitors comprising an ionic liquid-based polymer electrolyte and functionalized CNT fibre electrodes show capacitance (48 F g$^{-1}$) and energy density (16.5 Wh kg$^{-1}$), also superior to those of pristine CNT fibres.

## 2. EXPERIMENTAL SECTION

**2.1. Functionalization of carbon nanotubes:** CNT fibers have been produced *via* continuous direct spinning process from the gas-phase where CNTs growth occurs over floating catalyst chemical vapor deposition (CVD) at 1250 ºC in controlled H$_2$ atmosphere. Ferrocene, butanol and thiophene have been used as catalyst source, carbon source and promoter, respectively, at the fixed weight ratio chosen to produce few-layer multiwalled CNTs.[10] The fibers have been collected at a draw ratio of 6.3 on mylar plastic substrate overlapping filaments on each other. As-spun material has been densified through exposure to acetone forming CNT fiber film which can be easily detached from the substrate. The preformed 2x2 cm$^2$ films (4-6 µm) have been subjected to electrochemical functionalization in two-electrode system using stainless steel mesh as a counter electrode. CNT fibers have been treated by chronoamperometry technique applying different constant voltage (2.5V, 5V and 10V) for one minute in 0.1 M Na$_2$SO$_4$ aqueous solution. For furher characterization functionalized samples were washed in distillated water and dried at 90 ºC overnight.

**2.2. Characterizations**: Raman spectroscopy measurements have been performed using Renishaw microspectrometer (50) equipped with He−Ne laser at 532 nm. TEM micrographs have been obtained with FEG S/TEM (Talos F200X, FEI) microscope operating at 80 kV. Surface elemental analysis by means of X-ray photoelectron spectroscopy (XPS) have been carried with a SPECS GmbH spectrometer using a monochromatic Al K$_\alpha$ source (hv = 1486.71 eV).

Water droplet spreading experiment has been conducted with an optical microscope OLYMPUS BX51 equipped with a camera OLYMPUS ColorView. A small drop of a liquid (2 µL) was placed on the surface of CNT fiber film and the focused regions of interest have been recorded by a video camera under an optical microscope. The area of the droplet at time zero when contacted with the fiber surface and time-resolved wicking area covered by liquid have been extracted from the video (recorded at a speed of 6 frames per second), and further analyzed.

Small-angle X-ray scattering (SAXS) experiments have been performed at the Non Crystalline Diffraction (NCD) beamline 11, ALBA Synchrotron Light Facilities. A microfocus of 10-microns diameter and radiation length of 1 Å have been used. SAXS patterns have been collected on Pilatus 1M detector (Dectris) with an active image area of 168.7 x 179.4 mm$^2$ and total number of pixels 981 x 1043. SAXS patterns have been analyzed using DAWN Science software and radial profiles have been obtained in the range of the scattering vector, $q$, 0.01 < q (Å$^{-1}$) < 0.15 from full 360º integration after subtracting the background and having calibrated the sample holder position by using a reference material (silver behenate, AgBh). Structural analysis from SAXS data here included has taken into account density fluctuations and thus, structural parameter of pristine and exfoliated fibre have been corrected from this contribution.[11,12] The methodology to obtain pore structure descriptors in CNT fibre-based electrodes is fully described by Santos et al.[13]

Electrochemical properties of pristine and functionalized CNT fiber films have been evaluated by cyclic voltammetry (CV) and electrochemical impedance spectroscopy (EIS) using a Biologic VMP multichannel potentiostatic–galvanostatic system with an impedance module. CV tests with scan rate varied from 5 to 200 mV s$^{-1}$ and EIS measurements with frequency ranged from 100 mHz to 200 kHz have been carried out in 3 electrode cell configuration with platinum mesh as the counter electrode, Hg/HgO reference electrode and 1M Na$_2$SO$_4$ aqueous electrolyte. Specific capacitance was calculated by integrating the area under CV curves and normalizing by mass of active material using the following equation:

$$C_{sp} = \int i\, dV / {mv\Delta V} \qquad (1)$$

Symmetric SCs have been assembled in a two electrode Swagelok® cell using cellulose paper as separator and 1M KOH or 1M Na$_2$SO$_4$ as electrolytes and tested by galvanostatic charge-discharge with current density range from 0.5 to 20 mA cm$^{-2}$. All-solid SCs have been assembled using polymer electrolyte membranes consisting of Pyr$_{14}$TFSI ionic liquid and Poly(vinylidene fluoride-cohexafluoropropene) (PVDF-co-HFP). The fabrication of PE membranes and fabrication of free-standing all-solid EDLC devices have been performed according to the procedure described before.[14] Specific capacitance of full cells was obtained from the slope of discharge curve as $C_{cell}$ = I/slope. Specific capacitance of a single electrode in the symmetric device was obtained as $C_s$ = 4 $C_{cell}$. Values of real energy ($E_{real}$) and power ($P_{real}$) densities were calculated by integrating discharge curves of full devices according to the equations:

$$E_{real} = I \int V dt \qquad (2)$$

$$P_{real} = \frac{E_{real}}{t_{dis}} \qquad (3)$$

## 3. RESULTS AND DISCUSSION

The samples used in this work are free-standing electrodes consisting of a unidirectional nonwoven fabric of CNT fibers, produced by winding a continuous individual CNT fibre directly from the floating catalyst chemical vapour deposition reaction. Functionalization was carried out by applying a constant voltage in a two electrode system using CNT fibers as working electrode, stainless steel mesh as counter electrode in 0.1 M $Na_2SO_4$ aqueous electrolyte (Figure 1a). Upon applying a high voltage above 2 V, there is rapid gas evolution accompanied by extensive electrode swelling in a few seconds, as shown in Figure 1b-c for a treatment at 10V for 1 minute. This short time was selected as optimum; longer functionalization times produced excessive exfoliation, thus limiting the ability to handle samples as free-standing structures. A fast functionalization process is also interesting with the view of applying it as a continuous process coupled to fibre spinning. After the treatment there is substantial change in macroscopic morphology and newly-acquired hydrophilicity, with the sample clearly absorbing a large amount of water after withdrawal from the cell (Figure 1c).

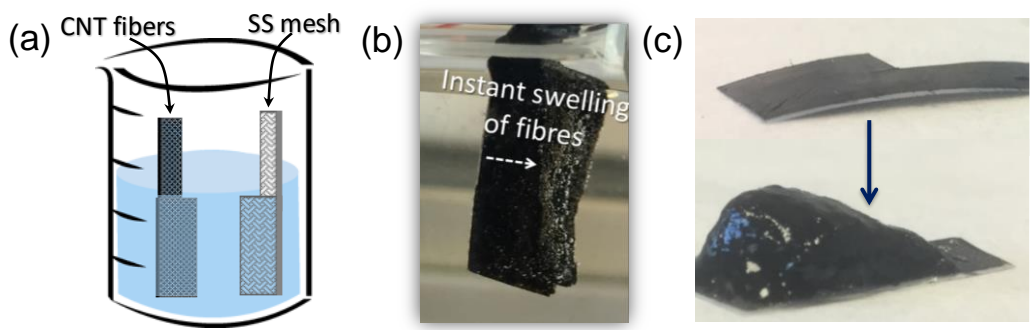

*Figure 1.a) Schematic of the electrochemical functionalization set up, b) a digital image of CNT fibers swollen during applying voltage. c) Digital images of a CNT fiber film before (top) and after (bottom) electrochemical functionalization.*

The interest then, is in understanding the different stages of the electrochemical functionalization process and in correlating treatment parameters to sample morphology, surface chemistry and electrochemical properties. For this purpose, samples functionalized at different voltages were washed in distilled water, dried at 80 °C and subjected to different characterization techniques.

Figure 2a depicts representative Raman spectra of the pristine and functionalized CNT fibres at different voltages. Each of them consists of three main signals, including disorder induced D-band at 1348 cm$^{-1}$, the tangential G-band at 1582 cm$^{-1}$, and the 2D-band at 2686 cm$^{-1}$ attributed to the two-phonon lattice vibrational process.[15,16] The spectrum of the pristine fibers reveals a low intensity of D-band and asymmetric lineshape of the G-band with a hint of a G$^-$ shoulder, which indicates that the CNTs are highly graphitic and have few layers.

After electrochemical modification of the CNT fibres the D-band intensity increases and the G band decreases, which confirms the introduction of defects in the CNTs that lower the crystallinity of their quasi-infinite conjugated lattice structure. The intensity ratio between D and G band ($I_D/I_G$) is often used as a convenient tool to monitor the relative degree of disorder in CNTs caused by functionalization.[17] The low initial value of 0.2 for the pristine material increases substantially (> 0.8) for all the voltages tested, peaking at around 1.3 (average) for 10 V. While this is indicative of a more defective material, i.e. successfully functionalized, these values are still relatively low, comparable to high-quality commercial multiwalled nanotubes, for example. This ensures that a high degree of graphitization is preserved in the material, which is beneficial to retain the superlative axial properties of the CNTs. More importantly, Raman spectra acquired from different layers through the thickness of the CNT fiber electrode showed values of $I_D/I_G$ within 0.9-1.5, confirming a relatively homogeneous distribution of functional groups introduced by the electrochemical processes, more uniform than that obtained by ozone functionalisation (ESI, Figure S1).

The introduction of functional groups is also manifested by the increase in the disorder-induced D' band at 1620 cm$^{-1}$, assigned to in-plane vibrations of the outer parts of graphite domains. Likewise, there is a gradual enlargement of the D+G-band at 2943 cm$^{-1}$ and the appearance of a small peak at 1140 cm$^{-1}$, commonly attributed to the formation of trans-(CH)$_n$ polyacetylene through cutting of CNT walls by oxygen radicals during oxygen evolution reaction at high voltages.[18]

In addition, the G and 2D bands of samples subjected to the electrochemical treatment have a blue upshift of about 3-6 cm$^{-1}$ and 2-8 cm$^{-1}$, respectively. This is most likely attributed to charge transfer induced by oxygen-containing functional groups acting as electron withdrawing moieties, similar to p-type dopants.[16] However, we note that electrochemical debundling of CNTs in a bundle could lead to similar effects.[15]

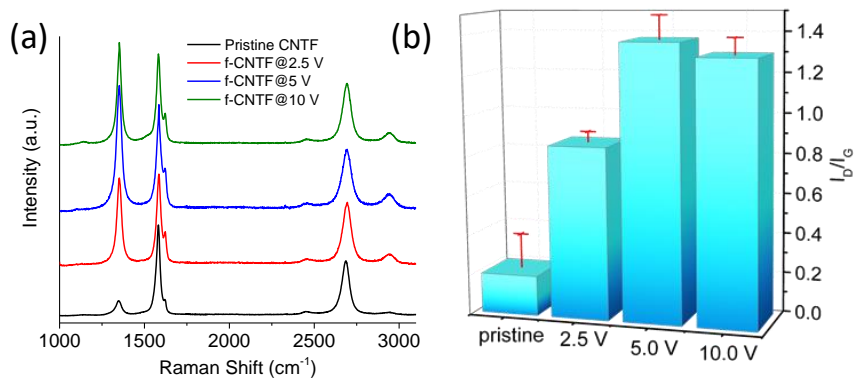

*Figure 2. a) Raman spectra of pristine CNT fibers and after functionalized at different voltages, b) evolution of the $I_D/I_G$ ratio upon increase of the voltage applied.*

The microstructural changes in the functionalized CNT fibers compared to the pristine material were observed under FESEM and TEM. As shown in Figure 3 and Figure S2,3 electrochemical functionalization resulted in significant alteration in the CNT structure through the formation of holes, broken layers, partial unzipping and debundling of CNTs. A comparison of high resolution TEM images of different samples corroborates that increasing bias voltage produces more extensive amorphization of the CNT structure. Whereas in the material treated at 2.5 V functionalization is mainly observed in outer CNT layers (Figure 3b,f) along with few instances of debundling, the functionalized CNT fibers at 5V retain the graphitic backbone but show extensive regions of defective CNT walls and debundled CNTs (Figure 3c,g and Figure S3i-l). Using a bias voltage of 10V produces greater removal of outer CNT layers and the formation of graphitic debris (Figure 3d, g and Figure S3m-p). Evidence of partial unzipping of CNTs is also observed at this high voltage (Figure 3h and Figure S3o). These results confirm that the bias voltage used has an effect of the type of defects formed. They also highlight the limitations of using Raman spectroscopy alone to characterize the introduction of functional groups, which is dominated by resonant modes that can mask out defects.

The effect of applied bias voltage on the evolution of new functional groups on CNT fibres was investigated using high resolution XPS. In the XPS survey scan of the samples, pronounced peaks corresponding to C1s and O1s transition were observed (Figure S4). Figures 4a and b present the representative C1s spectra of pristine and functionalized CNT fibers at 5V, respectively, which were calibrated by the C1s binding energy at 284.5 eV. It can be seen that the line shape of the C1s survey spectra after electrochemical treatment of the samples becomes broader and more asymmetric at the high energy binding side, which appears to be more pronounced upon increase of the voltage of functionalization (ESI, Figure S5). This suggests that the contribution of different bonding configurations of carbon based functionalities change with applied bias voltage. To estimate the relative population of carbon hybridizations a careful deconvolution of the C1s peaks has been performed.[19] The results in Figure 4a reveal presence of peaks for pristine CNT fibers corresponding to sp$^2$-hybridized C=C (284.4 eV), sp$^3$ C-C (284.9 eV), π-π* transitions at higher binding energy range (290.5 eV) and oxygen-containing species C-O (286.4 eV) mainly attributed to presence of amorphous carbonaceous impurities formed during CVD process.

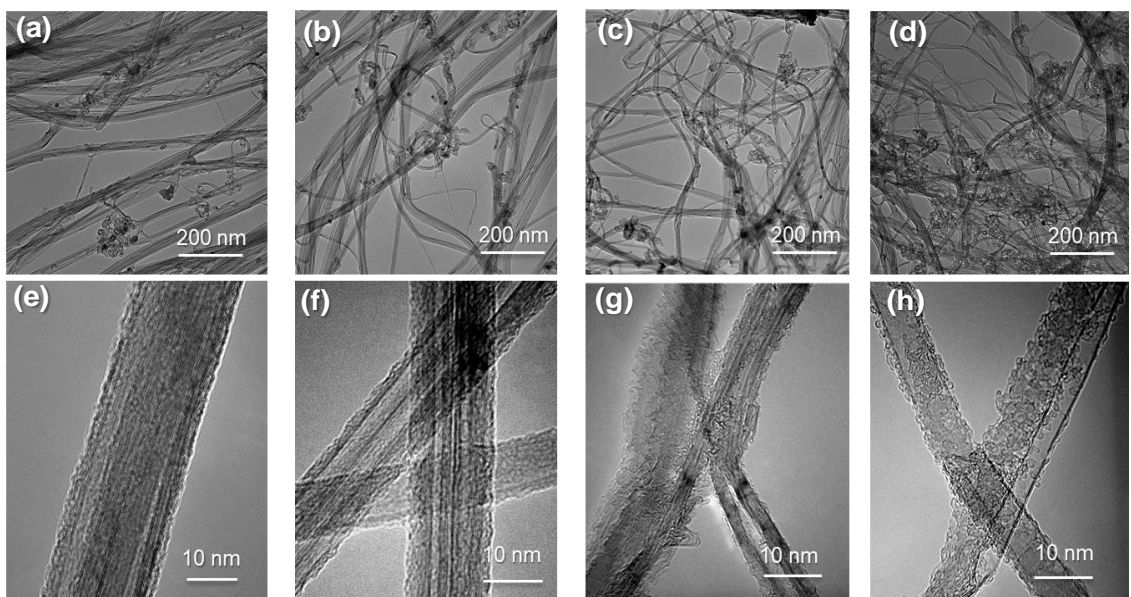

*Figure 3. Low magnification (a-d) and high magnification (e-h) TEM images of pristine (a,e) and functionalized CNT fibers at 2.5 V (b,f), 5 V (c,g), 10 V (d,h) showing reduction of CNTs graphitization degree and growth of amorphous carbon fraction at higher voltage.*

Deconvolution of the C1s peak for functionalized samples (Figures 4b and S5a,b) indicates the evolution of carbon population at high binding energies of 285-290 eV with increasing applied bias voltage, which is associated to introduction of new oxygen-containing functional groups, also accompanied by reduction of graphitic features. The bar chart in Figure 4c and Table S1 show the variation in the concentration of different oxygenated functional groups for pristine and functionalized CNT fibres. It is clear that the extent of oxygen incorporation (grey and yellow bars for C-O and C=O respectively in Figure 4c) and sp$^3$ C concentration (orange bars in Figure 4c) increases with the higher voltage applied during electrochemical treatment of CNT fibers, whereas the contribution of sp$^2$ carbon and π-π* transition band (green bars in Figure 4c) exhibits a gradual decay. Importantly, the aggressive electrochemical treatment at 10V bias leads to emergence of carboxylic O-C=O groups (289.9 eV) and complete disappearance of π-π* transition band. This can be related to dramatic modification of CNTs structure originated from debundling, unzipping of nanotubes and consequent increase of sidewalls defects, which is in a good agreement with TEM observations.

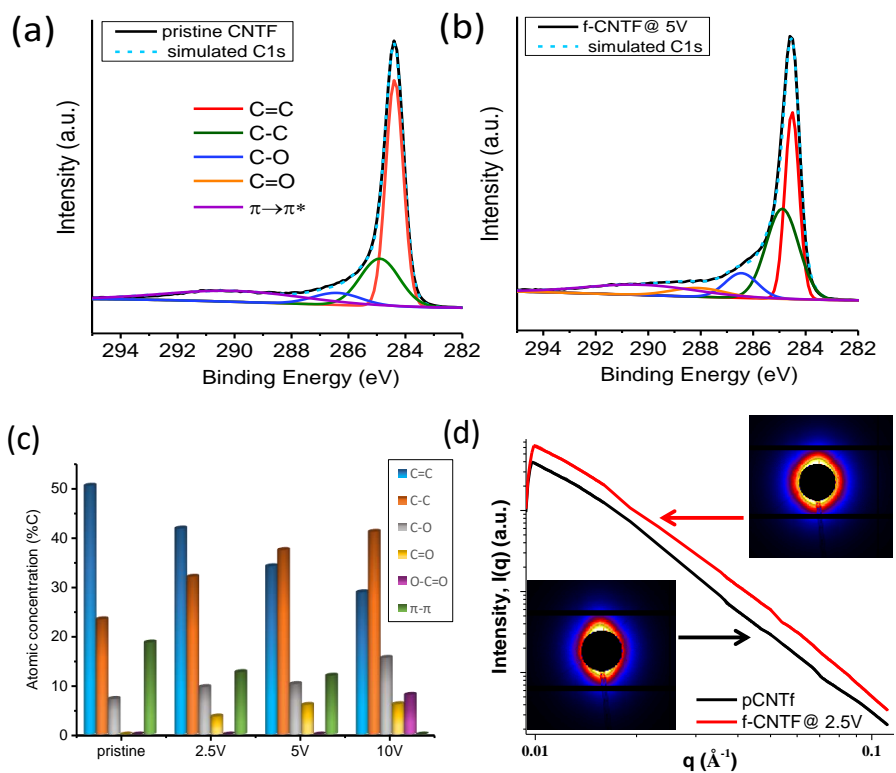

*Figure 4*. a) Normalized XPS spectra in the C1s region for pristine and b) 5V functionalized CNT fibers, and c) bar chart describing relative population of C=C. C-C, C-O, C=O, COO and π- π transition in CNTs. (d) Radial profile of SAXS from pristine CNT fiber (black) and 2.5V exfoliated CNT fibers (red). Insets show the modification in the diffuse scattering, thus in the porous structure of CNT fibres, due to the exfoliation process.

In order to gain further insight into the structural changes of the CNT fibre samples after exfoliation SAXS measurements were performed on pristine and exfoliated CNT fibers. SAXS can provide a direct indication of the surface-to-volume ratio as well as several descriptors of the porous structure of carbon-based materials[11,20]. A comparison of SAXS radial profiles from pristine CNT fibre and 2.5 V exfoliated CNT fibre (Figure 4d) shows an increase in scattering intensity throughout all $q$ values. This confirms the formation of new interface boundaries between CNT and air, that is, that turbostratic CNT/CNT interfaces are replaced with CNT/air interfaces, which must therefore originate from debundling of CNTs due to the electrochemical exfoliation.

Following an analysis method we recently proposed for CNT fibre-based electrodes,[13] the average bundle size extracted from SAXS measurements ($l_{bundle}$) is reduced from 38 to 10 Å after 2.5V exfoliation. In addition, the form factor of the CNTs[21,22], which is observed in the pristine as a small hump at $q \sim 0.08$ Å$^{-1}$, is not present in the exfoliated fibre. This can be taken as further indication of the modification of the fibre in terms of the size and shape of the bundles.[23]

Understanding the process of electrochemical functionalization, particularly, the formation of various oxygenated functional groups with applied voltage, requires consideration of the different interrelated processes taking place simultaneously during the treatment: (i) electrochemical wetting of pristine CNT fibres, (ii) oxidation of edges and outermost CNT layers and (iii) evolution of molecular oxygen and carbon dioxide.

Prior to the electrochemical treatment, the pristine CNT fibre (CNTF) electrode is hydrophobic and has a glossy mirror-like appearance when immersed in the aqueous electrolyte (Figure S6b), an effect arising from light interference at the solid/liquid/air interface similar to that in hydrophobic plant surfaces. When a bias voltage is applied, the hydrophobic surface of pristine CNTs undergoes instant wetting by the electrolyte at a bias of around 1.7 V and the electrodes appears opaque and black due to water infiltration (Figure S6b).[24] This ensures that functionalization occurs throughout the electrode inner pore structure.

Under an anodic bias voltage above the electrochemical stability window of water (1.23 V *vs.* RHE), $H_2O$ molecules undergo water splitting and evolution of oxygen (OER) will occur at the CNT fiber. At this positive voltage, oxidation of CNT will also take place starting preferably at the high energy sites, such as the edge planes, defects etc., as shown schematically in Figure 5.[25,26,27] This oxidation leads to the formation of hydroxyl (C-OH) carbonyl (C=O), epoxy (C-O-C), carboxylic acid (COO-) groups, along with oxidation of water to molecular oxygen.[28] Further oxidation of the oxygenated functional groups could cause evolution of CO ($E^0$=0.518 V) and $CO_2$ ($E^0$=0.207 V versus RHE).[29] As a result, huge swelling of CNT fibres as well as accumulation of gas bubbles inside the fiber structure were observed during functionalization (Figure 1b). Interestingly, the cyclic voltammogram acquired immediate after exfoliation of CNT fibres, shows presence of noticeable redox peaks which disappears after removal of bubbles from the exfoliated samples, followed by several potential cycling (ESI, Figure S7). These peaks might be attributed to labile oxygenated functional groups generated during exfoliation presenting faradaic contribution in aqueous electrolytes, as well as of electroreduction of $O_2$.

With increasing functionalization voltage, the kinetics of carbon oxidation reaction is accelerated, which increases the population of the oxygenated functional group, as observed from XPS analysis. Under higher bias voltage, loss of carbon content from the nanotube by possible evolution of CO and $CO_2$ results in oxidative cleavage,[29] which in turn leads to unzipping of carbon nanotubes (Figure 5b) as well as an increase in functionalization of the material, which is consistent with TEM observation of more defective surfaces and the presences of carbonaceous debris.

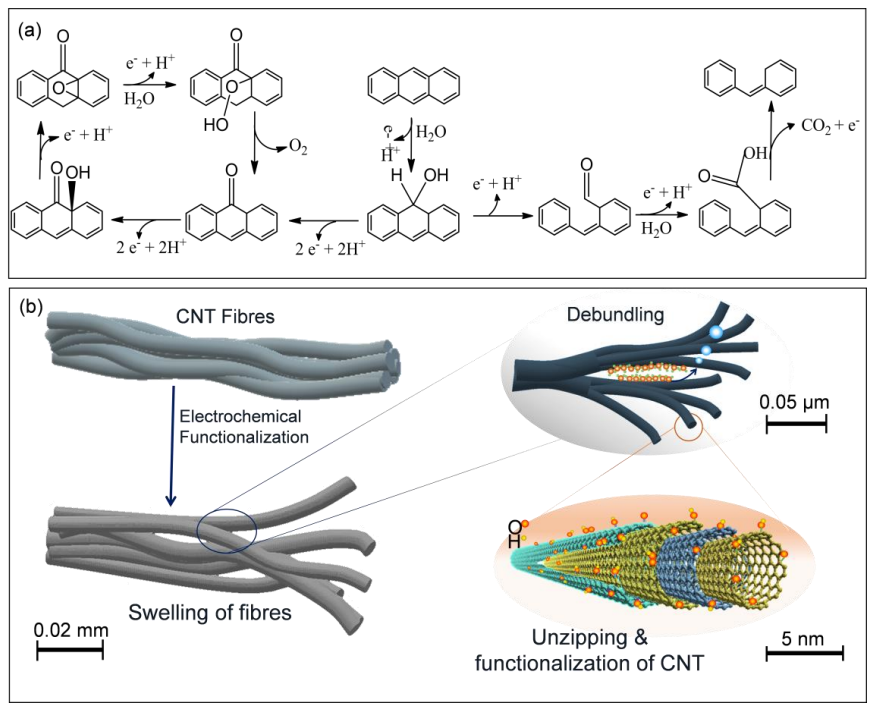

*Figure 5. (a) Scheme for plausible reactions during electrochemical functionalization showing generation of surface functional groups (C-OH, CHO, CO, COOH, CH$_2$, C-O-C) and evolution of molecular oxygen and carbon dioxide. (b) Schematic representation of the macroscopic changes in the CNT fibres during the functionalization showing swelling of CNT bundles, originated from the debundling, gas evolution, surface functionalization and unzipping of CNTs.*

In order to relate the effects of functionalization on the polarity of the functionalized CNT fiber films in detail, droplet wicking experiments were performed (Figure 6a-d. ESI, Figure S8 -14). Water, benzyl alcohol (BA) and toluene were chosen as representative liquids to follow the polar and non-polar nature of the CNT fiber surfaces. Figures 6e and 6g show the increase of wicking area with respect to time for 2 μL water and benzyl alcohol droplets, respectively. Due to extreme hydrophobicity of pristine CNTs, the wicking area of water droplet remains constant over initial time (until evaporation starts), whereas benzyl alcohol (and toluene) wicks and spreads rapidly. We analyze the dynamics of liquid spreading/wicking for CNT fibre fabrics before solvent evaporation by an expression of the form

$$\frac{S}{S_0}\left[\ln\frac{S}{S_0} - 1\right] = -1 + \frac{W}{S_0}t \qquad (i)$$

Here $S$ and $S_0$ are the wicking area at time $t$ and $0$ respectively. $W$ is a constant which modulates wicking rate according to porosity, interfacial energy (wetting) and liquid viscosity. For the case of liquid penetration into two parallel disks, the radial flow between the disks has been previously expressed as

$$\frac{S}{S_0}\left[ln\frac{S}{S_0}-1\right] = -1 + \frac{2\pi r\gamma cos\theta}{3\eta S_0}t \qquad (ii)$$

Where, $r$ is the capillary radius, $\eta$ the viscosity, $\gamma$ is the surface tension of the liquid and $\theta$ is the contact angle. Equation (ii) has been widely used to describe liquid spreading in thin porous macroscopic media, such as fabrics.[30] Its success lies in providing a simple method to measure the dependence of spreading rate on viscosity (ESI, Figure S11) and interfacial energy.

Equipped with this relation, Figures 6f, h and Figure S12 clearly show that with increasing functionalization voltage increases, the normalized wicking rate of polar water droplet increases (Figure 6e), while a reverse trend is observed for the less polar benzyl alcohol droplet (Figure 6g). The plots of ($S/S_0$ [ln $S/S_0$ -1]) *vs*. wicking *t* for different solvents give slopes that correlate with the expected chemical affinity or degree of wetting. Indeed, the calculated contact angle values capture the increase hydrophilicity upon electrochemical functionalization, as summarized in Table 1.

Further we estimated the dispersive and polar components of surface energy (*SE*) of these functionalized CNT fibres using the Owens- Wendt equation:[31]

$$\gamma_L\ cos\theta = -\gamma_L^D - \gamma_L^P + 2\sqrt{\gamma_S^D\gamma_L^D} + 2\sqrt{\gamma_S^P\gamma_L^P} \qquad (iii)$$

Here $\gamma_S^D$ and $\gamma_S^P$ are the respective dispersive and polar components of surface energy of the carbon nanotube surface, and $\gamma_L^D$ and $\gamma_L^P$ are the dispersive and polar components of the surface tension of the liquids respectively ($\gamma_S = \gamma_S^D + \gamma_S^P$). The polar and dispersive surface energies of the liquids used are included in Table ST2 in ESI. In Figure 7a the resultant surface energy values of the functionalized CNT surfaces are plotted against the corresponding functionalization voltage. The surface energy of pristine CNTF was estimated to be 37.4±1.6 mJ/m$^2$, which is in in agreement with the surface energy of CNT (42.1 mJ/m$^2$), reported by Kim et al.[32] and comparable to, though lower than that of graphene (46.7 mJ/m$^2$) estimated using the Newmann model by Wang *et al.*[33] Interestingly, the pristine material has around 5% polar contribution to the total surface energy, which can be attributed to presence of defects and edges as well as possibility of H⋯π interaction of the curved-π cloud with protic probe solvents.[34]

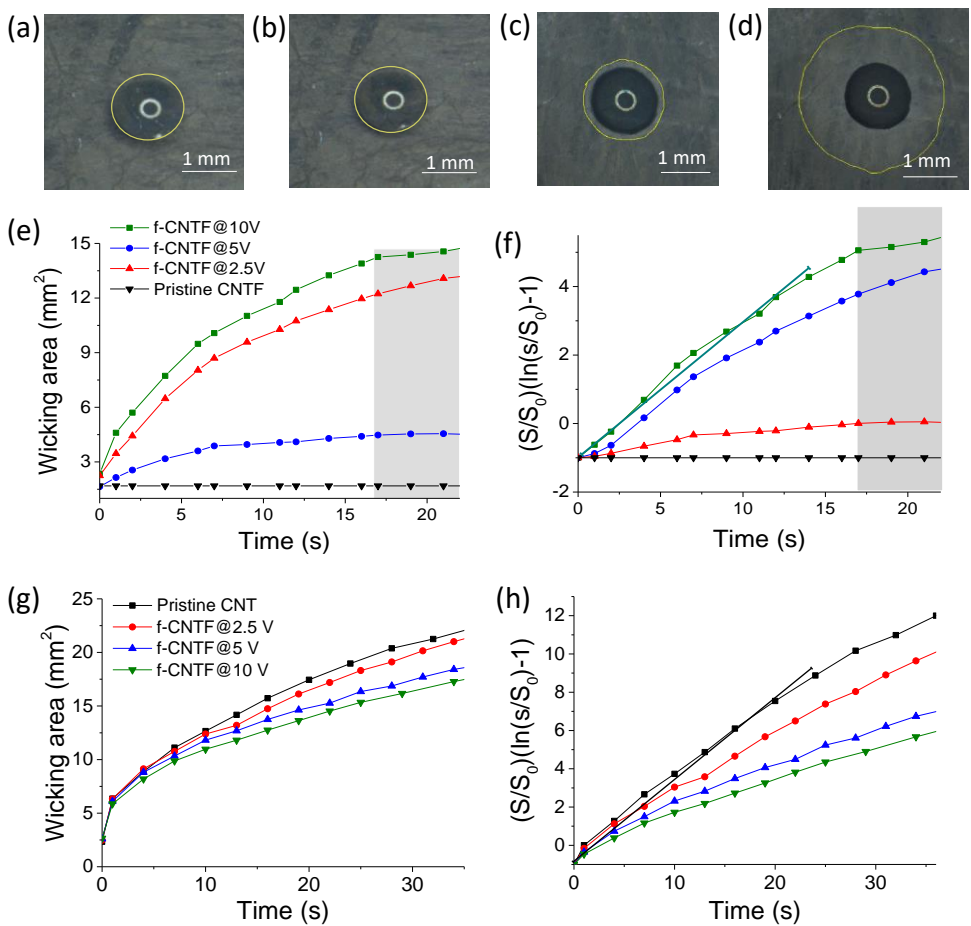

***Figure 6.*** *Digital images of water droplet on pristine CNTF film at (a) 1 s and (b) 2.5 min, and the same on functionalized CNTF film @5V at (c) 1s and (d) 20 s. Plots of wicking area (e, g) and $(S/S_0)(ln(S/S_0)-1)$ (f, h) of water droplet (e, f) and benzyl alcohol droplet (g, h) as a function of time. The grey region in figure e and f correspond to dominating evaporation kinetics, leading to non-linear wicking behavior.*

**Table 1**: Summary of estimated contact angle values of water, benzyl alcohol and toluene droplets on functionalized CNT fibres at different voltages.

|  | Contact angle of water droplet | Contact angle of benzyl alcohol droplet | Contact angle of toluene droplet |
|---|---|---|---|
| CNTF (pristine) | - | 64.4° | 21.3° |
| f-CNTF @2.5 V | 89.1° | 66.9° | 40.5° |
| f-CNTF @5.0 V | 83.9° | 69.6° | 49.3° |
| f-CNTF @10.0 V | 81.4° | 73.9° | 60.4° |

A drop of surface energy was observed for the functionalized CNTF compared the pristine one (21.6±3.7, 22.1±5.0 and 23.9±5.1 for f-CNTF@2.5V, @5V and @10V respectively), together with an increase in the contribution of polar SE with increasing functionalization voltage. We found that the polarity percentages (polar SE/ total SE) gradually increase from 6% for the pristine sample to 26, 48 and 69 % for the functionalized samples at 2.5, 5 and 10V bias voltage, respectively. This trend could be further related to the concentration of newly acquired oxygen-containing functional groups of CNT fibers determined by XPS. As shown in Figure 7b, the polarity fraction estimated from the Owens- Wendt equation correlates well with the weighted dipole moment of the materials (WDM), a product of the percentage of functional group population and corresponding dipole moment, obtained from XPS analysis. The dipole moment values for C-O, C=O and COO- are considered as 0.7, 2.5 (average) and 2.02 Debye, respectively. The plot of polarity % over WDM can be fitted with a straight line (red line in Figure 7b) with slope of 1.7. We attribute the fact that this slope is greater than one to the fact that the polarity values measured from liquid wetting kinetics include contributions from the chemical affinity of the surface towards the liquid as well as surface topology[35–37] which are not captured by the simple model used here.

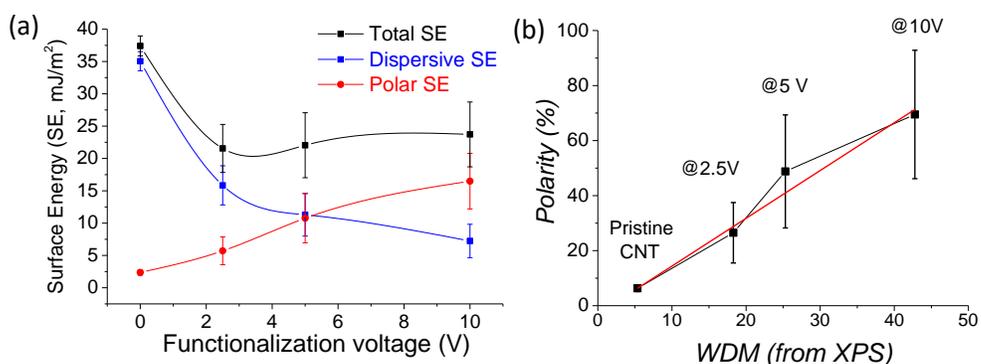

*Figure 7. (a) Plot of the dispersive and polar components of the surface energies with respect to functionalization voltage. and (b) polarity% with respect to weighted dipole moment (WDM, as obtained from XPS analysis). To estimate WDM, the dipole moment values used for C-O, C=O and COO are 0.7, 2.5 (averaged) and 2.02 respectively.*

The electrochemical properties of these functionalized CNT fibres with different degrees of hydrophilicity were studied by cyclic voltammetry (CV) in a three electrode cell in 1M $Na_2SO_4$ aqueous electrolyte medium. Figure 8a shows a typical rectangular-like shape of CVs for pristine films, implying pure electrical double-layer capacitive behavior and showing little penetration of the electrolyte. The CV curves of functionalized CNTs at 2.5 and 5.0 V show a much larger enclosed area and the emergence of broad pseudocapacitive peaks due to fast redox reactions of newly-acquired surface functional groups.[38]

The CV of sample functionalized at 10 V is totally distorted, losing capacitive behavior and showing a pronounced pair of peaks at 0.25 V and 0.55 V. This is probably due to the excessive oxidation of the CNT fibers resulting in widespread loss of sp$^2$ conjugation and instead formation of amorphous carbonaceous material, as observed by XPS. The shape of the CVs of the functionalized samples and their relative area compared to pristine CNT fibers were found to be reproducible for nominally identical experimental conditions, for example, when comparing samples functionalized in our laboratory with those functionalized at the synchrotron facility where in-situ SAXS measurements were performed.

The plot of capacitance against scan rate in Figure 8b shows the overall substantial improvement in specific capacitance of functionalized CNT fibers, achieving a maximum for 5V functionalization (60 F g$^{-1}$ at 5 mV s$^{-1}$), compared to the as-produced material (15 F g$^{-1}$ at 5 mV s$^{-1}$). Poor electrochemical behavior of the pristine material in aqueous electrolyte is attributed to the high hydrophobicity of CNT fibers translated into non-sufficient infiltration of the liquid into porous network of the electrodes. By contrast, electrochemical treatment of the carbon films causes considerable improvement of their wettability which increases substantially the effective surface area available for ion adsorption. Reversible redox processes of oxygen-containing groups on the surface of CNTs give rise to additional pseudocapacitance of the electrodes, but this effect is highly convoluted with the newly acquired hydrophilicity. The results also show that functionalization at 10V does not lead to further enhancement of electrochemical performance, on the contrary, it produces a decrease in capacitance due to extensive degradation of the material and a large reduction in electrical conductivity as indicated by impedance measurements (ESI, Figure S14a). This is also corroborated by the fast capacitance decay with scan rate in Figure 8b.

The performance of functionalized samples has been also characterized in symmetric supercapacitor devices assembled in two-electrode cells using 1M Na$_2$SO$_4$ and 3M KOH aqueous electrolytes. Figure 8c depicts the charge discharge (CD) profiles of the SCs based on pristine and 5V functionalized CNT fibers. Both devices with a similar mass loading of active material demonstrate nearly ideal triangular shape of CD curves with high coulombic efficiency >95%, which indicates excellent reversibility of cells. The capacitance values derived from the slope of CD profiles reveal about four fourfold increase of specific capacitance for 5V functionalized sample in comparison with pristine CNT fibers in 1M Na$_2$SO$_4$. Oxygenated groups do not exhibit pseudocapacitance in neutral electrolytes (similar to PYR$_{14}$TFSI) meaning that the fourfold increase is mainly due to enhanced wettability of functionalized CNT fibers in aqueous electrolytes (enhanced "effective surface area"). A sevenfold increase of specific capacitance was observed in 3M KOH electrolyte where it is well-known that pseudocapacitive contribution of oxygenated groups is much more pronounced than in neutral or aprotic electrolytes (ESI, Figure S14b).[39] Thus, this significant increase might be attributed to the combined effect of enhanced wettability and additional contribution of pseudocapacitance to the total capacitance of the materials in functionalized CNT. As a consequence, supercapacitors based on electrochemically modified electrodes also experiences a large increase in energy density (Figure 8d) from 0.11 to 0.58 Wh kg$^{-1}$ with $Na_2SO_4$ electrolyte and from 0.13 to 0.93 Wh kg$^{-1}$ using KOH. The specific power of the device calculated at different current densities remains in similar range of 0.4-12 kW kg$^{-1}$, which is consistent with unchanged conductivity of 5V functionalized fibers. For reference, the CNT fibre functionalized electrochemically at 5V was found to have comparable capacitance as that functionalized via a gas phase process by exposure to an ozone cleaner for 30 minutes, with both samples tested in alkaline medium.[7]

To further evaluate the potential of functionalized CNT fiber electrodes in real devices, symmetric all-solid supercapacitor devices were fabricated using a polymer electrolyte (PE) membrane based on $Pyr_{14}TFSI$ ionic liquid (IL) and PVDF-co-HFP copolymer, following a simple assembly method previously described by the group,[14] whereby all-solid SCs are produced by simply sandwiching and pressing together two CNT fiber electrodes and a PE membrane. The use of IL enables achieving full wetting of electrodes while providing ionic conductivity and a broad electrochemical window of 3.5 V. The comparison of CD profiles in Figure 9a indicates substantial improvement of electrochemical properties for the device assembled with @5V f-CNT fiber sheets compared to pristine electrodes. Specific capacitance increases from 33 to 48 F g$^{-1}$ at 1mA cm$^{-2}$ and energy density from 12 to 16.5 Wh kg$^{-1}$, respectively (Figure 9b, c). Note that the use of this hydrophobic polymer electrolyte excludes incomplete wetting of CNTs and mitigates contribution of faradaic redox reactions as a cause of the increase in capacitance. Instead, as shown previously, it is related to the electronic structure of the low-dimensional CNTs and due to the increase in quantum (chemical) capacitance upon introduction of oxygen-containing functionalities, effectively representing new energy states near the Fermi level.[7]

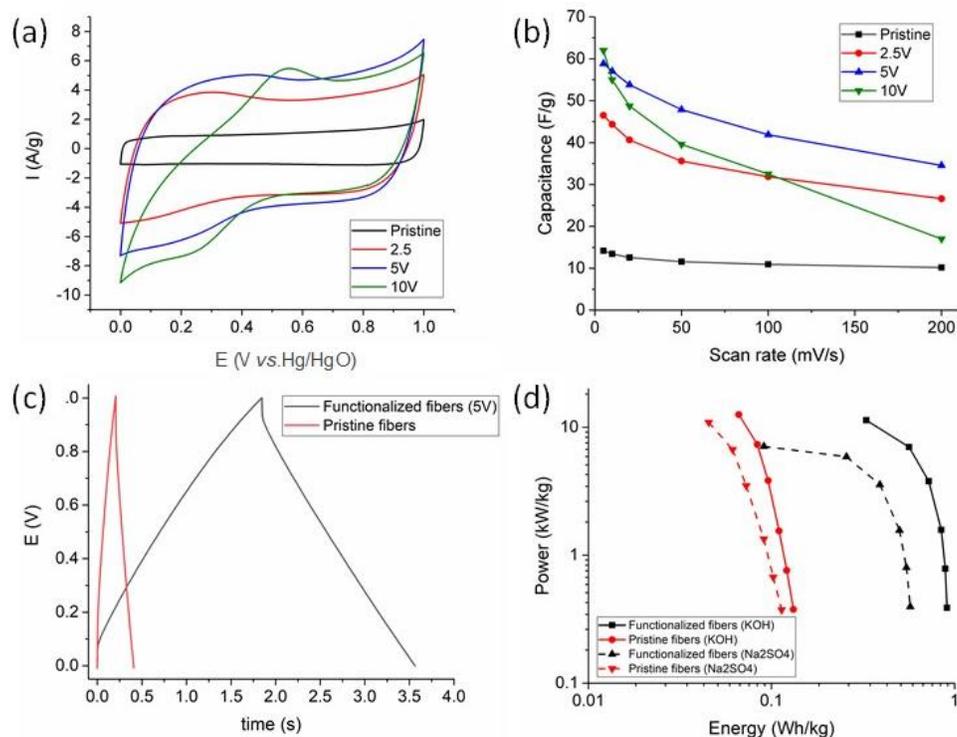

***Figure 8.*** *(a) CV curves at scan rate of 100 mV s$^{-1}$ and b) capacitance plots obtained in 1M Na$_2$SO$_4$ electrolyte for pristine and functionalized at different voltages CNT fibers; (c) CD profiles of pristine and functionalized CNT fibers measured at 2 mA cm$^{-2}$ in 1M Na$_2$SO$_4$ electrolyte and d) Ragone plot comparing electrochemical performance of pristine and 5V functionalized CNT fibers in symmetric supercapacitor device using different electrolytes.*

Finally, we have assembled a free-standing all-solid 3 cm$^2$ supercapacitor (SC) device employing @5V functionalized CNT fiber electrodes and tested its electrochemical stability under flexural conditions. Figure 9d illustrates CD profiles measured for the device at unbent state and under 180° deformation after 100 bending cycles. The discharge curves exhibit similar nearly ideal straight lines, indicating that repeating bending of SC does not induce any deterioration in its performance. Furthermore, flexural 180º deformation of the all-solid device gives rise to the enlargement of the area under discharge profile producing a 12% increase in energy density. Such behavior can be associated with improved infiltration of the polymer electrolyte into the porous media of the carbon electrodes, as has been previously observed for flexible EDLCs based on CNT fibers.[14] In addition, Figure 9d (inset) demonstrates an example of 3 cm$^2$ flexible all solid SC device assembled with @5V functionalized CNT fiber electrodes lighting red LED in a bent state.

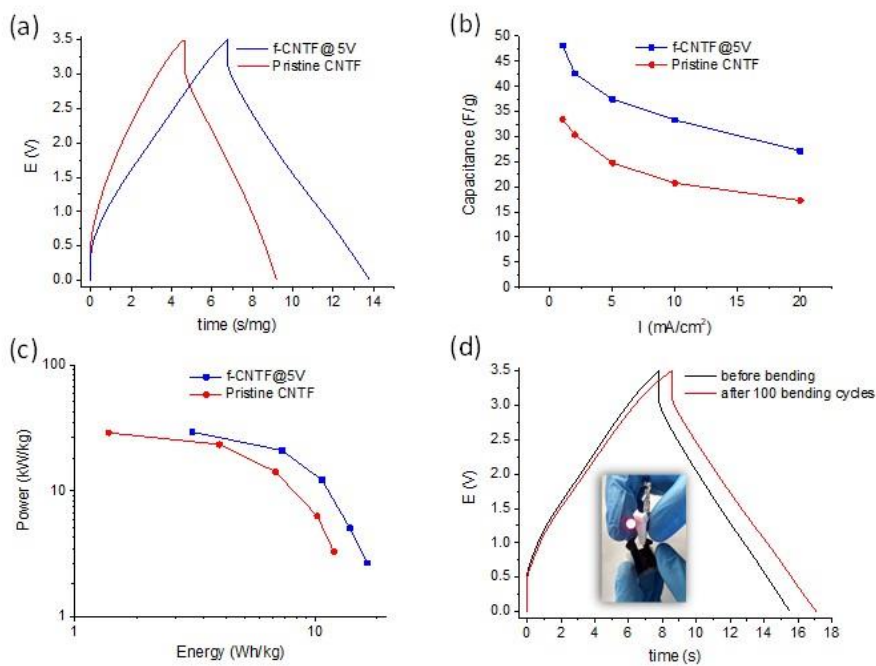

***Figure 9.*** *(a) CD profiles at current density of 5 mA cm$^{-2}$, b) specific capacitance and c) Ragone plots comparing electrochemical performance of pristine and 5V functionalized CNT fibers in symmetric all-solid SC device assembled with Pyr$_{14}$TFSI-based PE membrane; (d) CD curves at 2 mA cm$^{-2}$ obtained for free-standing all-solid 3 cm$^2$ SC device based on @5V f-CNT fibers before and after 100 bending cycles and (inset) an image of all-solid SC device lighting red LED in a bent state.*

## 4. CONCLUSIONS

In conclusion, this work introduces a fast and controlled electrochemical functionalization method to modify CNT fiber assemblies. By adjusting the voltage applied during electrochemical functionalization it is possible to change the degree of functionalization and thereby polarity of the surface. Based on Raman, XPS and TEM characterization, the optimum conditions of electrochemical treatment in terms of the balance between newly-acquired functionalities and preservation of initial morphology corresponds to functionalization at bias voltage of 5V.

By tracking the spreading of different liquids in thin CNT fibre samples we could estimate the polar and dispersive components of surface energy and confirm the increase in polarity of functionalized fibres with increasing voltage, with good agreement with dipole moment determined from XPS. This simple method to quantitatively determine changes in hydrophilicity and solvent affinity could be a very useful tool to characterize a wide range of porous nanocarbon materials.

The increased hydrophilicity of functionalized CNT fibers along with newly-acquired redox active functional groups results in large increase of specific capacitance and energy density of the electrodes in aqueous electrolytes and in an ionic liquid. This points to the fact that the changes in total capacitance in functionalized fibres are not only due to increased affinity with the electrolyte and the presence of pseudocapacitve redox processes, but also to increases in quantum (chemical) capacitance and possible changes in specific surface area. Separating the contribution from effects could provide further guidelines to optimize functionalization treatments by adjusting the concentration of functional groups at the CNT surface while preserving a highly conducting graphitized backbone in the electrode structure.

Finally, the benefits of this electrochemical functionalization method of carbon nanotubes are demonstrated by assembling a current collector-free all-solid flexible SC, which shows an increment in specific capacitance from 33 to 48 F g$^{-1}$ and energy density from 12 to 16.5 Wh kg$^{-1}$, as compared to pristine CNT fiber-based SC.

## AUTHOR INFORMATION

### Corresponding Author

\* rebeca.marcilla@imdea.org, juanjose.vilatela@imdea.org

## ACKNOWLEDGEMENT

Financial support is acknowledged from the European Union Seventh Framework Program under grant agreement 678565 (ERC-STEM) and Clean Sky-II 738085 (SORCERER JTI-CS2-2016-CFP03-LPA-02-11), and from MINECO (MAT2015-62584 ERC, RyC-2014-15115, Spain) Synchrotron XRD experiments were performed at NCD beamline at ALBA Synchrotron Light Facility with the collaboration of ALBA staff.